\documentclass[aps,twocolumn,floatfix]{revtex4}
\usepackage{epsfig}

\begin{document}

\title{Depinning transition of dislocation assemblies:
pileup and low-angle grain boundary}

\author{Paolo Moretti}
\affiliation{Dipartimento di Fisica, Universit\`a "La Sapienza",
P.le A. Moro 2, 00185 Roma, Italy} \affiliation{Center for
Materials Science and Engineering, University of Edinburgh, King's
Buildings, Kenneth Denbigh Building, Edinburgh EH93JL, UK}
\author{M.-Carmen Miguel}
\affiliation{Departament de F\'{\i}sica Fonamental,
Facultat de F\'{\i}sica, Universitat de Barcelona, Av. Diagonal 647,
E-08028, Barcelona, Spain}
\author{Michael Zaiser}
\affiliation{Center for Materials Science and Engineering,
University of Edinburgh,
King's Buildings, Sanderson Building, Edinburgh EH93JL, UK}
\author{Stefano Zapperi}
\affiliation{INFM UdR Roma 1 and SMC, Dipartimento di Fisica,
Universit\`a "La Sapienza", P.le A. Moro 2, 00185 Roma, Italy}

\begin{abstract}
We investigate the depinning transition occurring in dislocation
assemblies. In particular, we consider the cases of regularly
spaced pileups and low angle grain boundaries interacting with a
disordered stress landscape provided by solute atoms, or by other
immobile dislocations present in non-active slip systems. Using
linear elasticity, we compute the stress originated by small
deformations of these assemblies and the corresponding energy cost
in two and three dimensions. Contrary to the case of isolated
dislocation lines, which are usually approximated as elastic
strings with an effective line tension, the deformations of a
dislocation assembly cannot be described by local elastic
interactions with a constant tension or stiffness. A nonlocal
elastic kernel results as a consequence of long range interactions
between dislocations. In light of this result, we revise
statistical depinning theories and find novel results for Zener
pinning in grain growth. Finally, we discuss the scaling
properties of the dynamics of dislocation assemblies and compare
theoretical results with numerical simulations.
\end{abstract}
\maketitle

\section{introduction}

The depinning transition of individual dislocations gliding on
their slip plane has been widely investigated in the
past~\cite{NAB-82,BRA-86,IOF-87,SEV-91,DAN-97,ZAP-01} in order to
explain solid solution hardening \cite{NAB-79,BUT-93,NEU-93}, that
is, the increase of the yield stress value when solute atoms are
present in a crystal.  The presence of solute atoms changes the
local properties of the host material, resulting in a pinning
force on nearby dislocations \cite{BUT-93,NEU-93}. This is not the
only source of pinning, which can also be provided by particle
inclusions or by immobile dislocations in other inactive slip
systems \cite{NAB-79}. Several approximate calculations have been
performed in the past to obtain the depinning stress from a
statistical summation of individual pinning forces.  In this
respect, collective pinning theories have been very successful in
the case of diffuse weak pinning forces
\cite{LAB-70,LAB-69,LAR-79}, whereas the theory introduced by
Friedel \cite{Friedel} is appropriate in the case of localized
strong pinning centers.

>From the purely theoretical point of view, dislocations provide a
concrete example of a more general problem: that of driven elastic
manifolds in quenched random media \cite{KAR-98}. Apart from
dislocations, other examples of this general problem are domain
walls in ferromagnets \cite{LEM-98,ZAP-98}, flux lines in type II
superconductors \cite{BHA-93,SUR-99}, contact lines
\cite{ROL-98,SCH-00}, and crack fronts \cite{BOU-97,SCH-97}. In
recent years, a vast theoretical effort has been devoted to
understand the depinning transition as a non-equilibrium critical
phenomenon
\cite{KAR-98,NAT-92,LES-97,NAR-93,ERT-94,CHA-00,LED-02}.  The
morphology of a depinning manifold is generally found to be
self-affine and can be characterized by a roughness exponent.
Other scaling exponents have been introduced to characterize the
behavior of correlation lengths and times, the velocity above
depinning, and the avalanching motion observed as the critical
threshold is approached.  Quantitative predictions of the critical
exponents have been obtained analytically by the renormalization
group \cite{KAR-98,NAT-92,LES-97,NAR-93,ERT-94,CHA-00,LED-02}, and
have been confirmed by numerical simulations
\cite{LES-97,CUL-98,LAC-01,ROS-03,SCH-95,RAM-97,TAN-98,ROS-02}. In
the course of time, a deeper level of description and
understanding of this phenomenon has been achieved, going far
beyond a mere estimate of the depinning force, which has typically
been the original motivation to address the problem.

The analysis of the depinning transition in dislocation theory has
often been made in the approximation of a dislocation line tension
\cite{NAB-82,BRA-86,IOF-87,SEV-91,DAN-97}.  Hence, a dislocation
is considered to be a flexible string with a constant line
tension. This analogy is not fully accurate. In fact, the bending
of a dislocation produces long-range stress and strain fields
\cite{FOR-67,WIT-59} such that the self-energy of a dislocation
line segment depends on its interaction with all the other
segments on the line, and therefore depends of the particular
dislocation configuration. Long range interactions lead to a
wavevector-dependent effective line tension. This is also
well-known in the case of vortex lines in high temperature
superconductors~\cite{BLA-94} or, for instance, for a dislocation
line in a vortex crystal~\cite{MIG-97}. In spite of this
wavevector dependence, the main features of the depinning
transition are basically unchanged \cite{ZAP-01}, due to the fact
that this effective line tension is only varying logarithmically
with the wavevector. Numerical simulations indicate, however, a
slight difference in the roughness exponent in the two cases
which, at present, is still not well understood
\cite{ZAP-01,TAN-98}. A similar surface tension approximation is
also used in the theory of Zener pinning \cite{SMI-48,MAN-98}, to describe the
interactions between grain boundaries and solute atoms during
grain growth \cite{HAZ-91}.

\begin{figure}[t]
\centerline{\epsfig{file=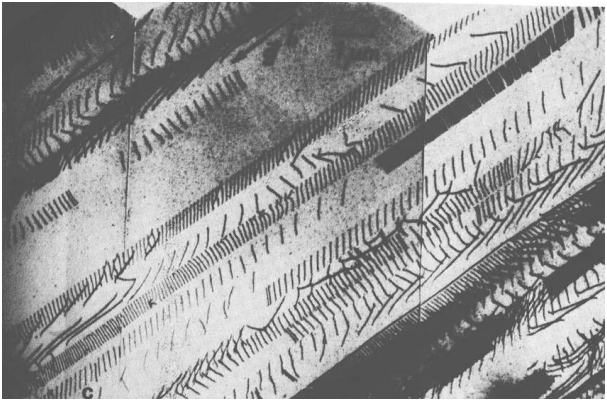,width=8cm,clip=!}}
\caption{Transmission electron micrograph taken from a
Cu-14.4at\%Al single crystal deformed at room temperature; the
image shows large regularly spaced dislocation pile-ups. Courtesy
of J. Plessing and H. Neuh\"auser~\cite{PLE-95}.}
\label{fig:pileup}
\end{figure}

While the behavior of an isolated dislocation pushed through a
random distribution of obstacles is at present quite well
understood, the results do not necessarily carry over to the more
realistic case of collective dislocation motion.  Dislocations
interact via their long-range stress fields, which may induce
intriguing jamming and avalanche-like phenomena even in the
absence of quenched pinning centers~\cite{MIG-02}. In most cases,
one can not simply neglect interactions and treat dislocations as
isolated objects.  The depinning transition of interacting
dislocation lines and/or loops of generic orientations and Burgers
vectors in a random solute distribution seems to be a formidable
task for theoretical treatment. In several instances, however,
dislocations are arranged into regular structures that are
amenable to a theoretical treatment.  In particular, here we
analyze the depinning transition of regularly spaced pileups and
low angle grain boundaries (LAGB). These relatively simple
structures are sometimes observed experimentally (see Fig.
~\ref{fig:pileup}) and provide a nice illustration of the effect of
dislocation interactions on the depinning transition. An early
analysis of the depinning of a dislocation pileup was presented in
Ref.~\cite{ZAI-79}, considering explicitly the emission of
dislocations from a source.

Here we address the problem by first computing the stress and elastic
energy associated to a small deformation of the dislocation
structure. This allows us to assess the validity of a local tension
approximation, which turns out to be completely inadequate in this
context. Long-range interactions between the dislocations involved
give rise to much larger self-stresses than for isolated dislocations.
As a result, pileups and low angle grain boundaries are much stiffer
than single dislocations. The elastic energy is then used in the
framework of statistical pinning theories to estimate the depinning
stress. In particular we consider collective pinning theory and
Friedel statistics. As an application of these results, we revise the
theory of Zener pinning in grain growth which originally made use of a
surface tension approximation for the elastic self-stress.  The
correct treatment of long-range stresses and dislocation interactions
inside the grain boundary yields a different result for the dependence
of grain size on material parameters.

At stresses close to the depinning stress, the dynamics of the
dislocation structures we are investigating exhibits critical
behavior which can be characterized in terms of scaling exponents.
Using previous renormalization group results, we gain a complete
quantitative picture of the depinning transition. In the elastic
approximation, pileups and low angle grain boundaries are
equivalent to a standard interface depinning problem with
long-range elasticity.  In two dimensions, the problem can be
mapped to a contact line or to a planar crack, which have been
extensively studied in the literature.  In three dimensions, the
self-stress is similar to the dipolar force in magnetic domain
walls and leads to logarithmically rough deformations.  In more
technical terms, $d=3$ is the upper critical dimension for the
transition, which is well described by mean-field exponents, up to
logarithmic corrections.

The scaling exponents associated with the depinning transition
describe not only the morphology of the dislocation assembly but also
its dynamics.  In order to confirm the validity of the elastic
calculations, on which we based the mapping with elastic manifolds, we
perform a series of numerical simulations for a dislocation pileup. We
consider a two dimensional system, neglecting the deformation of
single dislocations, which amounts to an effective one dimensional
particle model. Simulations of the model display results in agreement
with the theory and allow to illustrate some interesting dynamical
effects.  In particular, the pileup displays a zero temperature power
law creep relaxation which can be interpreted by scaling
relations. Below threshold, the power law relaxation terminates into a
pinned configuration, while above threshold there is a crossover to
linear creep or average constant velocity sliding.  As it is common
for this class of systems, the motion of the pileup takes places in
the form of avalanches whose distribution again can be characterized
by scaling exponents.

\bigskip
\section{Elasticity}

\begin{figure}[ht]
\centerline{\psfig{file=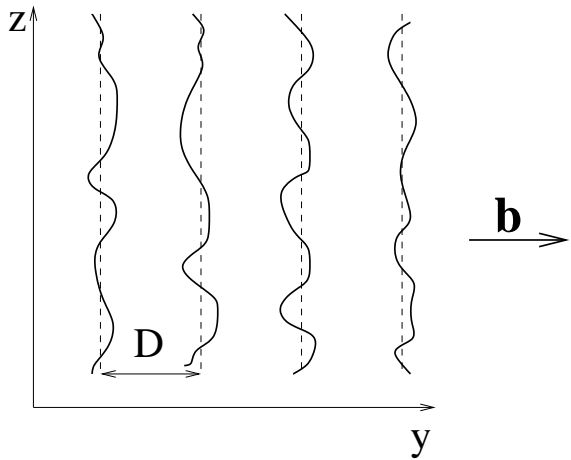,width=7.5cm,clip=!}}
\caption{A regularly spaced dislocation pileup with Burgers vector
along the $y$ axis. The ideal configuration is plotted with straight
dashed lines, whereas the solid lines represent their possible glide
deformations within the slip plane $yz$.}
\label{fig:1}
\end{figure}

Developing a theory for collective dislocation depinning requires
the basic knowledge of the elastic properties of the dislocation
assembly in the first place. In this section, we determine the
elastic response of two particular dislocation assemblies: a
regularly spaced pileup and a low angle grain boundary of edge
dislocation lines. The two structures are quite similar
geometrically; they are both one-dimensional arrangements of $N$
dislocation lines with the same Burgers vector {\bf b} and average
line direction $\hat{e}$ (for edge dislocations $\hat{e}\perp {\bf
b}$) along a given direction of space $\hat{d}$, but they differ
in the relative orientation of the Burgers vector and the
arrangement direction $\hat{d}$.  In particular, in a pileup a set
of edge dislocations lies in the slip plane, defined by the
dislocations axis $\hat{e}$ and the Burgers vector, so that
$\hat{d}\parallel \hat{b}$ (see Fig.~\ref{fig:1} for a particular
example with $\hat{e}=\hat{z}$ and $\hat{d}\parallel
\hat{b}\parallel \hat{y}$), while in the LAGB the edge
dislocations lie in the perpendicular plane such that
$\hat{d} \perp \hat{b}$ (see Fig.~\ref{fig:2} for a particular
geometry). Neglecting climb, i.e. the motion of a dislocation
perpendicular to its slip plane, deformations of the structure
occur solely within the slip plane both for the pileup and for the
LAGB.  In this section we derive the shear stress and the elastic
energy associated with small deformations of these dislocation
assemblies. This is needed in order to derive the yield stress
from statistical pinning theories. For completeness, we consider
the problem both in two and in three dimensions.

\subsection{Two dimensions}

In this particular case, we do not consider elastic deformations
along the dislocation line direction. This approximation is
relevant for thin crystals and amounts to treat the dislocations
in the structure as rigid lines, rendering a two-dimensional
problem. On the other hand, we take into account small variations
in the position of the dislocations in the one-dimensional
structures they form. We consider the case of LAGB and then
directly extend the result to the pileup case. In fact, in linear
approximation the elastic energy turns out to be the same in both
cases.

\begin{figure}[ht]
\centerline{\psfig{file=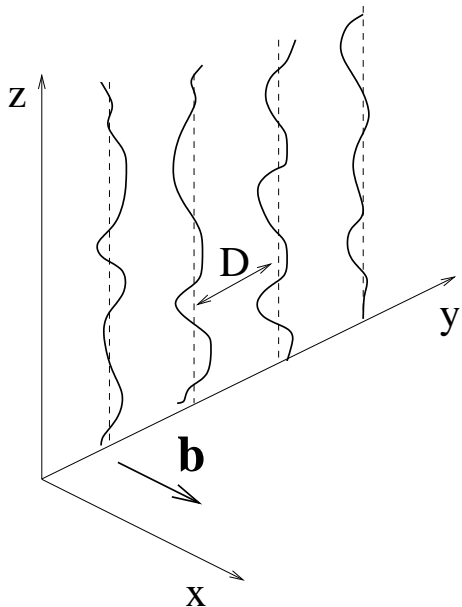,width=6cm,clip=!}}
\caption{A regularly spaced low angle grain boundary where the
dislocations Burgers vector is parallel to the $x$ axis. The ideal
configuration is plotted with straight dashed lines in the plane $yz$,
whereas the solid lines represent their possible glide deformations
within the slip plane $xz$.}
\label{fig:2}
\end{figure}

Here and throughout the paper, we consider an ideal LAGB as an
infinite set of equally spaced edge dislocations lying on the $yz$
plane (without loss of generality we consider the plane $x=0$) with
Burgers vector pointing along the positive $x$ axis ${\bf b}=b\hat{x}$
(see Fig.~\ref{fig:2}). In the rigid dislocation approximation, we do
not consider the deformations along the $z$ axis so that each
dislocation is described by a set of coordinates $(x_n,y_n)$, where
$y_n=n D$, $D$ is the grain boundary spacing, and where $x_n$ is a
small displacement from the $x=0$ plane.  The shear stress field at a
given point $(x,y)$ due to a dislocation at $(x_n,y_n)$ is given by
\cite{Friedel,HIR-68}
\begin{equation}\label{sigmaxy}
\sigma_{xy}^n(x,y)=\frac{\mu b}{2\pi
(1-\nu)}\,\frac{(x-x_n)[(x-x_n)^2-(y-y_n)^2]}
{[(x-x_n)^2+(y-y_n)^2]^2}
\end{equation}
where $\mu$ is the shear modulus and $\nu$ is the Poisson ratio.  The
glide component of the total force per unit length on another
dislocation $m$ in the LAGB can be readily obtained from the
Peach-Koehler expression ${\bf f}=({\bf \sigma}\cdot {\bf b})\times
\hat{e}$~\cite{Friedel,HIR-68}
\begin{equation}
f_x(x_m,y_m)=b\sum_{n=-\infty}^{+\infty}\sigma_{xy}^n(x_m,y_m).
\end{equation}
For small deformations $|x_m-x_n|\ll D|m-n|$ we have
\begin{equation}
f_x(x_m,y_m)= -\frac{\mu b^2}{2\pi
(1-\nu)}\sum_{n=-\infty}^{+\infty}\frac{x_m-x_n} {(y_m-y_n)^2},
\end{equation}
which can be used to obtain the elastic energy
\begin{widetext}
\begin{equation}
E=-\sum_{m=-\infty}^{+\infty}\,\int
f_x(x_m,y_m)\,dx_m=\frac{\mu b^2}{8\pi
(1-\nu)}\sum_{m=-\infty}^{+\infty}\sum_{n=-\infty}^{+\infty}
\frac{(x_m-x_n)^2}{(m-n)^2D^2}\,\mbox{with}\,\,m\neq n.
\label{eq:en2d}
\end{equation}
\end{widetext}
It is instructive to express the elastic energy in Fourier space,
where one can easily identify the energy cost of the different
modes. For an infinitely long LAGB $N\rightarrow \infty$, we can write
the dislocation displacements as
\begin{equation}
x_m=\int_{BZ}\frac{dk}{2\pi} e^{-ikDm}\,x(k),
\end{equation}
where the integral is restricted to the first Brillouin zone (BZ) of
the reciprocal space, and using the following results
\begin{equation}
\sum_{d=1}^{+\infty}\frac{1}{d^2}=\frac{\pi^2}{6}, \mbox{\ \ \ }\,
\sum_{d=1}^{+\infty}\frac{cos(\gamma d)}{d^2}=\frac{\pi^2}{6}\,-
\,\frac{\pi |\gamma|}{2}\,+\,\frac{\gamma^2}{4}
\end{equation}
we obtain  \cite{chui}
\begin{equation}
E=\frac{\mu b^2}{8\pi (1-\nu)D^2}\int_{BZ}\frac{dk}{2\pi}\left
(2\pi |k|-Dk^2\right)\tilde{x}(k)\tilde{x}(-k).
\end{equation}
From this expression, one can see that the elastic interaction
kernel $(2\pi |k|-Dk^2)$ is not quadratic in the wavevector, as it
would be the case for a local elastic manifold with a constant
tension or stiffness, but grows roughly as $|k|$ for long
wavelength deformations. This is a consequence of long range
interactions between dislocations in the LAGB which render a much
stiffer structure. In the following sections we will explore the
relevant consequences of this result in the collective depinning
of such dislocation structures, something that has been
disregarded in previous dislocation depinning studies.

The elastic energy associated with perturbations of a regularly
spaced dislocation pileup can be obtained in an analogous manner.
According to the geometric conditions assumed here, all Burgers
vectors are now oriented along the positive $y$ axis, and since
the dislocations are all in the same slip plane we can now write
$x=x_n=0$. Proceeding as before, the total Peach-Koehler force on
dislocation $m$ along the new glide direction is given by
\begin{equation}
f_y(0,y_m)=\frac{\mu b^2}{2\pi (1-\nu)}\sum_{n=-\infty}^{+\infty}\frac{1}{y_m-y_n}.
\end{equation}
Notice that the Peach-Koehler force is now repulsive, however the
stability of the system is ensured in the case of an infinite pileup
where the dislocations located at the extremes (at $\pm \infty$) have
fixed positions, or equivalently, for a finite pileup with periodic
boundary conditions. Thus one can also compute the elastic energy cost
of small displacements $\delta y_m$ of the dislocations in the pileup
with respect to their stable positions. Up to first order in $\delta
y_m$, we obtain a recovery elastic force
\begin{equation}
f_y(0,\delta y_m)=-\frac{\mu b^2}{2\pi
(1-\nu)}\,\sum_{n=-\infty}^{+\infty} \frac{\delta y_m-\delta
y_n}{(y_m-y_n)^2}
\end{equation}
equivalent to the one obtained for the case of the LAGB. The
corresponding elastic energy cost
\begin{equation}
\tilde{E}=\frac{\mu b^2}{8\pi (1-\nu)D^2}\int_{BZ}\frac{dk}{2\pi}\left
(2\pi |k|-Dk^2\right)\delta y(k)\delta y(-k)
\end{equation}
has the same form with a long-range interaction kernel. Thus as a
partial conclusion of this section, we can emphasize that long
wavelength distortions of low angle grain boundaries and equally
spaced pileups of straight dislocation lines with translational
invariance along the dislocation axis have the same nonlocal elastic
properties, with eigenvalues that grow linearly with the modulus of
the wavevector considered.

\subsection{Three dimensions}

In this section, we consider the more general and realistic case
of deformable dislocation lines where, according to the geometry
established in the previous section, we loose the translational
invariance along the $z$ axis. As before, we consider first the
case of a LAGB with Burgers vectors oriented along the $x$ axis,
in which each dislocation can be described by a set of coordinates
$(x_n(z),y_n,z)$, where again $y_n=n D$, but where now the
displacement $x_n(z)$ with respect to the $x=0$ plane depends on
the height of the infinitesimal dislocation segment considered
(see Fig.~\ref{fig:2}). The elastic stress field due to a general
dislocation line or loop can be obtained, for instance, by
considering the line as being composed of elementary segments of
infinitesimal length~\cite{HIR-68}. Depending on the relative
orientation of the Burgers vector and the local tangent vector
$\hat{\tau}(z)$, each segment can either have edge ($\hat{\tau}(z)
\perp {\bf b}$) or screw character ($\hat{\tau}(z)
\parallel {\bf b}$), or it can be a combination of both. A first
approximation of a general dislocation line can be its
parametrization in terms of a succession of only edge and screw
segments ~\cite{Footnote1}. The mathematical form of the elastic
stress fields generated by these two types of elementary segments
is simpler and renders amenable the analytic treatment of the
problem. For instance, the shear stress field due to an edge
dislocation segment of length $\Delta z'$ and located at the point
$(x',y',z')$ is given by~\cite{HIR-68}
\begin{equation}\label{ss1e}
\sigma_{xy}(x,y,z)=\frac{\mu b}{4\pi (1-\nu)}\,\frac{x-x'}{R_0^3}
\left[1-3\,\frac{(y-y')^2}{R_0^2}\right]\Delta z',
\end{equation}
where
\begin{equation}
R_0^2=(x-x')^2+(y-y')^2+(z-z')^2.
\end{equation}
On the other hand, the shear stress field due to an screw segment of
length $\Delta x'$ is~\cite{HIR-68}
\begin{equation}\label{ss1s}
\sigma_{xy}(x,y,z)=-\,\frac{\mu
b}{4\pi}\,\frac{z-z'}{R_0^3}\,\Delta x'.
\end{equation}

Equations (\ref{ss1e}) and (\ref{ss1s}) allow us to calculate the
glide component of the total Peach-Koehler force ${\bf f}=({\bf
\sigma}\cdot {\bf b})\times \hat{\tau}$ on an edge or an screw
segment. The glide force on an edge segment at $(x_m(z),y_m,z)$ has
two components due to any other edge $f_x^{EE}$ or screw segments
$f_x^{SE}$
\begin{eqnarray}
f_x^{EE}(x_m(z),y_m,z)&=&\frac{\mu b^2}{4\pi
(1-\nu)}\,\frac{x_m(z)-x_n(z')}{R_{mn}^3(z,z')}
\nonumber\\
&&\left[1-3\,\frac{(y_m-y_n)^2}{R_{mn}^2(z,z')}\right]\Delta z'\,\Delta z,
\\\nonumber
f_x^{SE}(x_m(z),y_m,z)&=&-\,\frac{\mu
b^2}{4\pi}\,\frac{z-z'}{R_{mn}^3(z,z')}\,\frac{\partial
 x_n(z')} {\partial z'}\Delta z'\,\Delta z,
\end{eqnarray}
respectively. Note that up to first order in the small
displacements ($x_m(z)-x_n(z')\simeq 0$), the relative distance
among segments can be written as
$R_{mn}^2(z,z')=(y_m-y_n)^2+(z-z')^2$. On the other hand, from the
general expression for the Peach-Koehler written above, it is
straightforward to verify that there are no glide forces acting
upon any screw segment on the dislocation line. After summing up
all nonvanishing contributions, we can obtain the elastic energy
as for the two dimensional case (see Eq.(\ref{eq:en2d})). The
elastic energy can be expressed as the sum $E = E^{EE} + E^{SE}$
of the interaction energies between edge-edge and edge-screw
segments. These interaction energies are given by
\begin{widetext}
\begin{equation}
E^{EE}=-\frac{\mu b^2}{32\pi (1-\nu)}\sum_{m,n}\int\int dzdz'
\left[1-3\,\frac{(y_m-y_n)^2}{R_{mn}^2(z,z')}\right]\frac{[x_m(z)-x_n(z')]^2}{R_{mn}^3(z,z')},
\end{equation}
\begin{equation}
E^{SE}=\frac{\mu b^2}{16\pi}\sum_{m,n}\int\int dzdz'\,
\frac{z-z'}{R_{mn}^3(z,z')}\,x_m(z)\,\partial_{z'}x_n(z').
\end{equation}
\end{widetext}

As we did for the rigid line case, we can also express this
elastic energy in Fourier space in order to diagonalize the
interaction matrix and to obtain the wavevector dependence of the
interaction kernel between the different deformation modes. The
detailed calculation is rather lengthy, so we merely indicate the
procedure followed and the final results obtained. We evaluate
separately the energy contribution due to the self-interaction
between the constituent segments of each individual dislocation
line, i.e. $n=m$, which we denote by $E_0$, and the energy
contributions due to the interaction of dislocation segments lying
on different lines, i.e. $n\neq m$, which we refer to as $E_1$.
Proceeding this way, we find that the total energy is
$E=E_0^{EE}+E_0^{ES}+E_1^{EE}+E_1^{ES}$. We express the
dislocation displacements in terms of their Fourier modes,
\begin{equation}
x_m(z)=\int_{BZ}\frac{dk}{2\pi}\int \frac{dq}{2\pi}e^{-ikDm}e^{-iqz}\,x(k,q)
\end{equation}
and evaluate the self-interaction contributions for long
wavelength deformations $qa\ll 1$ where $a$ is a short-distance
cutoff introduced to preclude the interaction of a line segment
with itself. The result can be written as
\begin{widetext}
\begin{eqnarray}
E_0^{EE}&=&\frac{\mu b^2}{16\pi
(1-\nu)}\int_{BZ}\frac{dk}{2\pi}\int\frac{dq}{2\pi}
\frac{1}{D}\left[2\left(\gamma
-\frac{3}{2}+\mbox{ln}\,a|q|\,\right)q^2-\frac{a^2}{12}q^4\right]x(k,q)x(-k,-q)
\label{eq:e0}\\
E_0^{SE}&=&\frac{\mu b^2}{8\pi}\int_{BZ}\frac{dk}{2\pi}\int
\frac{dq}{2\pi} \frac{1}{D}\left[2(-\gamma
-\mbox{ln}\,a|q|\,)q^2+\frac{a^2}{2}q^4\right]x(k,q)x(-k,-q)
\end{eqnarray}
\end{widetext}
where $\gamma$ is the Euler constant. The usual quadratic
wavevector dependence of a local interaction kernel is modified in
this particular case by logarithmic corrections. This is known to
be the result of long range interactions in the case of a
dislocation line, as well as for similar singularities such as
vortex lines in high temperature superconductors~\cite{BLA-94}.

Similarly, the energy contributions due to interactions between
segments of different dislocation lines ($n\neq m$) in the LAGB can be
expressed as
\begin{widetext}
\begin{eqnarray}
E_1^{EE}&=&\frac{\mu b^2}{16\pi
(1-\nu)}\int_{BZ}\frac{dk}{2\pi}\int\frac{dq}{2\pi}
\frac{1}{D}\left[2\left(\gamma
+\mbox{ln}\,\frac{D|q|}{4\pi}\right)k^2+\frac{2\pi}{D}\frac{k^2}{(k^2+q^2)^{1/2}}
+\frac{D^2}{2\pi^2}\zeta(3)k^2q^2\right]x(k,q)x(-k,-q) \\
E_1^{SE}&=&\frac{\mu
b^2}{16\pi}\int_{BZ}\frac{dk}{2\pi}\int\frac{dq}{2\pi}
\frac{1}{D}\left[2\left(\gamma
+\mbox{ln}\,\frac{D|q|}{4\pi}\right)q^2+\frac{2\pi}{D}\frac{q^2}{(k^2+q^2)^{1/2}}
+\frac{D^2}{2\pi^2}\zeta(3)q^4\right]x(k,q)x(-k,-q),
\label{eq:e1}
\end{eqnarray}
\end{widetext}
where $\zeta(x)$ is the Riemann {\em zeta} function. Naturally,
the interaction kernel between the deformation modes for the three
dimensional grain boundary case depends explicitly on both the $y$
and $z$ components of the wavevector in an intricate manner.
Nevertheless, as in the two dimensional case, for long wavelength
deformations the leading term of the interaction kernel is
essentially linear in the wavevector, which manifests the
nonlocality of the interactions.

Finally, let us consider the case of a three dimensional pileup
with the same geometric specifications as adopted in the
rigid-line case (see Fig.~\ref{fig:1}). Since the Burgers vector
is oriented along the $y$ axis, the glide component of the
Peach-Koehler force on screw segments (now parallel to the $y$
axis) vanishes. On the other hand, there is a non-vanishing glide
force on edge segments due to the shear stress field generated by
any other segments. For instance, the shear stress field due to an
edge dislocation segment of length $\Delta z'$ and located at the
point $(x',y',z')$ in this case is given by~\cite{HIR-68}
\begin{equation}\label{ss1e_p}
\sigma_{xy}(x,y,z)=-\frac{\mu b}{4\pi (1-\nu)}\,\frac{y-y'}{R_0^3}
\left[1-3\,\frac{(x-x')^2}{R_0^2}\right]\Delta z'.
\end{equation}
Moreover, the shear stress generated by an screw segment of length
$\Delta y'$ is~\cite{HIR-68}
\begin{equation}\label{ss1s_p}
\sigma_{xy}(x,y,z)=\frac{\mu
b}{4\pi}\,\frac{z-z'}{R_0^3}\,\Delta y'.
\end{equation}

Without loss of generality we consider a pileup lying on the $x=0$
plane, and we account for small perturbations affecting the $y_n$
coordinates of the dislocation lines in the ideal assembly. Thus, we
can replace $y_m-y_n\,\rightarrow\,(y_m+\delta y_m(z)) -(y_n+\delta
y_n(z'))$, where the displacements depend on the $z$ coordinate of the
infinitesimal line segment considered. Expanding up to first order in
$(\delta y_m(z)-\delta y_n(z'))$, the resulting Peach-Koehler glide
forces on an edge segment due to other edge $f_y^{EE}$ or screw
segments $f_y^{SE}$ are
\begin{eqnarray}
&&f_y^{EE}(0,y_m+\delta y_m(z))=\nonumber\\
&&\frac{\mu b^2}{4\pi (1-\nu)}\,\frac{\delta y_m(z)-\delta
y_n(z')}{R_{mn}^3(z,z')}
\left[1-3\,\frac{(y_m-y_n)^2}{R_{mn}^2(z,z')}\right]\Delta
z'\,\Delta z,
\nonumber\\
&&f_y^{SE}(0,y_m+\delta y_m(z))=\nonumber\\
&&-\,\frac{\mu b^2}{4\pi}\,\frac{z-z'}
{R_{mn}^3(z,z')}\,\frac{\partial \delta y_n(z')} {\partial
z'}\Delta z'\,\Delta z,
\end{eqnarray}
where
\begin{equation}
R_{mn}^2(z,z')=(y_m-y_n)^2+(z-z')^2.
\end{equation}

The corresponding elastic energy is
\begin{widetext}
\begin{eqnarray}
E^{EE}&=&-\frac{\mu b^2}{32\pi (1-\nu)}\sum_{m,n}\int\int dzdz'
\left[1-3\,\frac{(y_m-y_n)^2}{R_{mn}^2(z,z')}\right]\frac{[\delta
y_m(z)-\delta y_n(z')]^2}{R_{mn}^3(z,z')},\\
E^{SE}&=&+\frac{\mu b^2}{16\pi}\sum_{m,n}\int\int dzdz'\frac{z-z'}
{R_{mn}^3(z,z')}\,\delta y_m(z)\,\partial_{z'}\delta y_n(z').
\end{eqnarray}\end{widetext}
Proceeding along the same lines indicated in the case of a LAGB,
i.e. separating the intra-line interactions from the inter-line
ones, it can be expressed in terms of Fourier modes as
$E=E_0^{EE}+E_0^{ES}+E_1^{EE}+E_1^{ES}$. The expressions
turn out to be equal to the one obtained for the LAGB and can be recovered
replacing $x(\pm k, \pm q)$ with $\delta y(\pm k, \pm q)$ in Eqs.~\ref{eq:e0}-\ref{eq:e1}.

%\begin{eqnarray}
%E_0^{EE}&=&\frac{\mu b^2}{16\pi
%(1-\nu)}\int_{BZ}\frac{dk}{2\pi}\int\frac{dq}{2\pi}
%\frac{1}{D}\left[2\left(\gamma
%-\frac{3}{2}+\mbox{ln}\,a|q|\,\right)
%q^2-\frac{a^2}{12}q^4\right]\delta y(k,q)\delta y(-k,-q)\\
%E_0^{SE}&=&\frac{\mu b^2}{8\pi}\int_{BZ}\frac{dk}{2\pi}\int
%\frac{dq}{2\pi} \frac{1}{D}\left[2(-\gamma
%-\mbox{ln}\,a|q|\,)q^2+\frac{a^2}{2}q^4\right]\delta y(k,q)\delta
%y(-k,-q)\\
%E_1^{EE}&=&\frac{\mu b^2}{16\pi
%(1-\nu)}\int_{BZ}\frac{dk}{2\pi}\int\frac{dq}{2\pi}
%\frac{1}{D}\left[2\left(\gamma
%+\mbox{ln}\,\frac{D|q|}{4\pi}\right)k^2+\frac{2\pi}{D}\frac{k^2}{(k^2+q^2)^{1/2}}
%+\frac{D^2}{2\pi^2}\zeta(3)k^2q^2\right]\delta y(k,q)\delta y(-k,-q)\\
%E_1^{SE}&=&\frac{\mu
%b^2}{16\pi}\int_{BZ}\frac{dk}{2\pi}\int\frac{dq}{2\pi}
%\frac{1}{D}\left[2\left(\gamma
%+\mbox{ln}\,\frac{D|q|}{4\pi}\right)q^2+\frac{2\pi}{D}\frac{q^2}{(k^2+q^2)^{1/2}}
%+\frac{D^2}{2\pi^2}\zeta(3)q^4\right]\delta y(k,q)\delta y(-k,-q).
%\end{eqnarray}
%

Thus also in this case we find wavevector dependent interaction kernels whose
leading terms (for low wavelength deformations) grow either
linearly with the wavevector in the case of interline
interactions, or quadratically with logarithmic corrections in the
case of intraline interactions. We thus can conclude that this
particular form of the elastic kernels is characteristic of the
long range interactions between different dislocations. 
As we will see in the following, the long-range elastic properties preclude
the analysis of the depinning transition of dislocation assemblies
based upon a local theory with a constant effective stiffness.

\section{Disorder: depinning transition}

Distortions in a LAGB or a pileup arise from interactions of the
dislocations with various kinds of impurities such as solute
atoms, precipitates or other immobile defects. The interactions
between individual dislocations and impurities have been computed
and are reported in the literature. For the purpose of this
article, we will consider quenched disorder created by a random
distribution of immobile impurities with concentration $c$ which
interact with dislocations via a force $f_p(r)=f_0 g(r/\xi_p)$,
where $f_0$ is the pinning strength, $\xi_p$ is the interaction
range and $r$ is the distance between the impurity and the
dislocation. The detailed shape $g(x)$ of the individual pinning
force is inessential for most purposes.

The morphology and dynamics of a pileup or a LAGB result from a
complicated interplay between elasticity and disorder. Pileup and
LAGB are examples of the general problem of the depinning of
elastic manifolds in random media, which has been extensively
studied in the past \cite{KAR-98}. In the elastic approximation,
the dynamics of the pileup or the LAGB follows
\begin{equation}
\chi\frac{\partial u}{\partial t}=\int d^dx' K(x-x')(u(x')-u(x))+b\sigma+\eta(x,u),
\end{equation}
where $\chi$ is a damping constant, $\sigma$ is the applied stress,
$\eta(x,u)$ describes the effect of the pinning centers and the
elastic interaction kernel $K$, computed in the previous section,
scales as $|k|$ in Fourier space. In the following we will discuss how
the main theoretical approaches to the depinning transition can be
applied to the problem at hand.

\subsection{Collective pinning theory: weak pinning}

Collective pinning theory describes the behavior of an elastic
manifold in the limit of weak disorder, when pinning is due to the
fluctuations of the random forces. The key concept is the introduction
of a characteristic length $L_c$ above which pinning becomes effective
(or energetically advantageous) and consequently the manifold is
distorted. The collective pinning length can be evaluated, for
instance, by balancing the elastic energy cost and the pinning energy
gain associated with a small displacement of a region of linear size
$L$. On scales below $L$, the manifold remains essentially undeformed
and, hence, the fluctuations in potential energy follow Poissonian
statistics. The effective concentration of the pinning defects along
the LAGB is given by
\begin{eqnarray}
\bar{c}_{\rm eff} = \left \{\begin{array}{ll}\bar{c}\quad\;,\;\xi_p>D\\
\bar{c}\frac{\xi_p}{D}\;,\;\xi_p<D \end{array}\right. \;{\rm
(2D)}\;\nonumber\\
c_{\rm eff} = \left \{\begin{array}{ll}c\quad\;,\;\xi_p>D\\
c \frac{\xi_p}{D}\;,\;\xi_p<D \end{array}\right. ;{\rm (3D)}\;.
\end{eqnarray}
The first expression refers to pinning by columnar defects of
areal concentration $\bar{c}$ in $d=2$, and the second to pinning
by localized defects of volume concentration $c$ in $d=3$. In d=2,
the characteristic energy of a section of a LAGB of size $L$
displaced by an amount of the order of $u$ can be written as
\begin{equation}
\bar{E}=\frac{\mu b^2 u^2}{D^2} -\bar{f}_0\xi_p\sqrt{\bar{c}_{\rm
eff} Lu}.
\end{equation}
Here both $\bar{E}$ and $\bar{f}_0$ are defined as quantities per unit
length. In the case of a thin film of thickness $h$, one can obtain
their three-dimensional counterparts just as $E=h\bar{E}$ and
$f_0=h\bar{f}_0$. Note the scale-independence of the nonlocal
expression of the elastic energy $\mu b^2 u^2/D^2$ in contrast to what
would be this energy in the local approximation $\mu b^2 u^2/DL$.
Essentially the same expression holds for the pileup. Balancing
elastic and pinning contributions and imposing that the displacement
is of the order of the pinning range $u\sim \xi_p$, one readily
obtains $L_c=(\mu^2b^4\xi_{\rm p})/(D^4 \bar{f}_0^2 \bar{c}_{\rm
eff})$. The LAGB is depinned when the work done by the external stress
in moving a segment of length $L_c$ over the distance $\xi_p$ exceeds
the characteristic pinning energy $\bar{E}(L_c)$ of this
segment. Equating $\bar{E}(L_c)=\sigma_c b L_c\xi_p/D$, for the case
above the result is given by $\sigma_c b =(\bar{c}_{\rm eff}
\bar{f}_0^2D^3)/(\mu b^2)$.

A similar calculation in $d=3$ is more subtle, since the elastic
and the pinning energies scale with the same power of $L$ and thus
cancel in the simple dimensional approach discussed above. As we
will discuss in the next section, this reflects the fact that $d=3$
is the upper critical dimension for the transition. To obtain
$L_c$ in this case, one should perform a perturbation expansion in
the disorder, as discussed in Ref. \cite{LAR-79} in the context of
the flux line lattice. One essentially computes the typical
displacement $u$ for a system of size $|{\bf r}|=L$, which for a
LAGB is given by

\bigskip\begin{eqnarray} \langle|u({\bf r}
)-u(0)|^2\rangle=\int\frac{d^2k}{(2\pi)^2}\int\frac{d^2k'}{(2\pi)^2}\,
(1-\mbox{cos}\,{\bf
kr}) \nonumber\\
G({\bf k})G({\bf k'})\,F({\bf k})F({\bf k'})
\end{eqnarray}
where $G({\bf k})$ is the Green function associated to the elastic
kernel determined in the previous section, and $F({\bf k})$ is the
pinning force density. In the spirit of collective pinning theory
$\langle F({\bf k})F({\bf k'})\rangle=W\delta^{(2)}({\bf k}+{\bf k'})$
with $W=(f_0\sqrt{c_{\rm eff}\xi_p})^2$. The explicit calculation
leads to the characteristic displacement

\begin{equation}
u(L)\simeq f_0\sqrt{c_{\rm eff}\xi_p}\,\frac{D^2}{\mu
b^2}\,\ln^{1/2}\frac{L}{D}
\end{equation}

This expression can then be inverted, imposing $u\sim\xi_p$, to obtain
\begin{equation}
L_c=D \exp \left[\frac{\xi_p}{c_{\rm eff}}\left(\frac{\mu b^2
}{f_0 D^2}\right)^2\right].
\end{equation}
The depinning stress can then be obtained as in $d=2$ and is given
by $\sigma_c b =(\mu b^2 \xi_p) /(D L_c)$. Again these results
generalize directly to the pileup case. It is however important to
note that they refer to the continuum limit, when one can neglect
the discrete nature of the dislocation system. To be consistent
with this assumption one should have $L_c \gg D$.

\subsection{Strong pinning: Friedel statistics}

Collective pinning is due to a statistical superposition of the forces
created by many obstacles. In the limit of strong and/or diluted
pinning centers, however, the characteristic bulge of width $\xi_p$
and extension $L_c$ as envisaged in the previous section may not
interact with enough pinning centres for this viewpoint to be
valid. Simple estimates for the boundaries of the collective pinning
regime are given by the inequalities $L_c \xi_p \ge 1/\bar{c}_{\rm eff}$ and
$L_c^2 \xi_p \ge 1/c_{\rm eff}$ for the d=2 and d=3 cases discussed above,
respectively.

In the regime of strong pinning, dislocations are pinned by individual
obstacles. The spacing of obstacles along the dislocation and the
depinning stress can be obtained by an argument which was, in the
context of single dislocations, developed by Friedel. The basic idea
is to consider the behavior of a dislocation segment as it depins from
a pair of strong obstacles. The length of the segment is $L$, and it
forms a bulge of width $u$. If the dislocation segment overcomes one
of the pins it will travel by an amount which is, again, of the order
of $u$ and, hence, sweep an area of the order of $L u$. Now we can
estimate the depinning threshold by requiring that during this process
the freed dislocation segment encounters, on average, precisely one
new obstacle. In other words, precisely at the point of depinning the
dislocation starts to move through a sequence of statistically
equivalent configurations. For a dislocation this leads to the
condition $Lu \simeq 1/(c\xi_p)$. $L$ and $u$ can be related by
equating the work done by the external stress $\sigma$ in bulging out
the dislocation to the concomitant elastic energy increase, $\Gamma
u^2/L = \sigma b u L$, where $\Gamma$ is a constant line
tension. Finally, the depinning force can be obtained by comparing the
external force $b\sigma L$ with the pinning force $f_0$. Solving these
three equations, one obtains the Friedel length $L_f\simeq
(\Gamma/c\xi_pf_0)^{1/2}$ and the depinning stress $\sigma_c b \simeq
(c\xi_pf_0^3/\Gamma)^{1/2}$.

This argument can be generalized in a straightforward manner to
the case of grain boundaries. Let us first consider the depinning
of a two dimensional LAGB as discussed above in the weak pinning
limit: In this case, the Friedel condition reads $Lu \simeq
1/\bar{c}_{\rm eff}$, the elastic energy per unit length of a bulge of
width $u$ and extension $L$ is $\mu b^2 u^2/D^2$ which must equal
the work per unit length $\sigma b L u/D$, and the force balance
(again per unit length) is $\sigma_c b L/D = \bar{f}_0$. Combining
these relations we find that the Friedel length and the depinning
stress are
\begin{equation}
L_f \simeq \frac{\mu b^2}{D^2 \bar{c}_{\rm eff} \bar{f}_0}~~~~\sigma_c
b \simeq \frac{\bar{c}_{\rm eff}\bar{f}_0^2 D^3}{\mu b^2} \quad \rm{(2
d)}.
\end{equation}
In 3d the Friedel condition is $L^2 u \simeq 1/c_{\rm eff}$, the
energy balance reads $\mu b^2 u^2 L/D^2 = \sigma b L^2 u/D$, and
the force balance is $\sigma_c b L^2/D = f_0$. This yields
\begin{equation}
L_f \simeq \frac{\mu b^2}{D^2 c_{\rm eff} f_0}~~~~\sigma_c b \simeq
\frac{c_{\rm eff}^2 f_0^{3} D^5}{\mu^2 b^4} \quad \rm{(3 d)}.
\end{equation}
Table 1 presents a compilation of results for the weak and strong
pinning cases in two and three dimensions. For comparison we have
also included results obtained under the assumption that the
elastic behavior of the grain boundary can, in local elasticity
approximation, be described by a scale-independent surface energy
$\Gamma_0 \sim \mu b^2/D$.

\begin{table*}
\begin{tabular}{| l | l | l | l | l | l | l |}
\hline model dimension & type of elasticity &
type of pinning & pinning length & critical stress & Zener exponent\\
\hline \hline 2d & local & weak & $L_c = \left(\frac{\Gamma_0^2
\xi_p}{\bar{f}_0^2 \bar{c}_{\rm eff}}\right)^{1/3}$ & $\sigma_c b
= \left(\frac{D^3\bar{f}_0^4 \xi_p \bar{c}_{\rm
eff}^2}{\Gamma_0}\right)^{1/3}$ & $m = 2/3$\\
\hline 2d & local & strong & $L_f =
\left(\frac{\Gamma_0}{\bar{f}_0 \bar{c}_{\rm eff}}\right)^{1/2}$&
$\sigma_c b = \left(\frac{D^2\bar{f}_0^3 \bar{c}_{\rm
eff}}{\Gamma_0}\right)^{1/2}$ & $m = 1/2$ \\
\hline 2d & non-local & weak & $L_c = \frac{\Gamma_0^2 \xi_p}{D^2
\bar{f}_0^2 \bar{c}_{\rm eff}}$ & $\sigma_c b = \frac{D^2
\bar{f}_0^2 \bar{c}_{\rm eff}}{\Gamma_0}$ & $m = 1$  \\
\hline 2d & non-local & strong & $L_f = \frac{\Gamma_0}{D \bar{f}_0 \bar{c}_{\rm eff}}$&
$\sigma_c b = \frac{D^2 \bar{f}_0^2 \bar{c}_{\rm eff}}{\Gamma_0}$& $m=1$\\
\hline 3d & local & weak & $L_c = \left(\frac{\Gamma_0^2
\xi_p}{f_0^2 c_{\rm eff}}\right)^{1/2}$& $\sigma_c b = \frac{D
f_0^2 c_{\rm eff}}{\Gamma_0}$ & $m=1$\\
\hline 3d & local & strong & $L_f = \left(\frac{\Gamma_0}{f_0
c_{\rm eff}}\right)^{1/2}$& $\sigma_c b = \frac{D f_0^2 c_{\rm
eff}}{\Gamma_0}$ & $m =1$\\
\hline 3d & non-local & weak & $L_c = D \exp
\left[\frac{\Gamma_0^2 \xi_p}{D^2 f_0^2 c_{\rm eff}}\right]$&
$\sigma_c b = \frac{\Gamma_0 \xi_p}{D} \exp
\left[-\frac{\Gamma_0^2 \xi_p}{D^2 f_0^2 c_{\rm eff}}\right]$& exponential \\
\hline 3d & non-local & strong & $L_f = \frac{\Gamma_0}{D f_0
c_{\rm eff}}$ & $\sigma_c b = \frac{D^3 f_0^3 c_{\rm
eff}^2}{\Gamma_0^2}$& $m=2$\\
\hline
\end{tabular}
\caption{Overview of pinning stresses and pinning lengths obtained
from different models}
\end{table*}

\subsection{Zener pinning in grain growth}

>From an experimental viewpoint, grain boundary pinning is
important since the mobility of grain boundaries may be a limiting
factor in grain growth \cite{SMI-48,MAN-98}. 
The dependence of the average grain size $R$ on the impurity concentration
$c$ goes under the name of Zener formula and was originally proposed
in Ref.~\cite{SMI-48} and generalized afterwards \cite{MAN-98}
\begin{equation}
R \propto c^{-m},
\end{equation}
where $m$ is the Zener exponent. Several theoretical and numerical 
derivations of this formula have been discussed in the literature,
often based on the local elasticity approximation \cite{MAN-98,HAZ-91}. 

Grain growth is driven by a reduction in energy: For an average grain
size $R$ and straight grain boundaries, the characteristic energy
stored per unit volume in the form of GB dislocations is of the order
of $\Gamma_0/R$ and, hence, the energy gain achieved by increasing the
grain size by d$R$ is $(\Gamma_0/R^2)$ d$R$.
Physically, the removal of GB dislocations occurs through the
motion of junction points in the GB network. As junction points
must drag the connecting GB with them, which may be pinned by
disorder, motion can only occur if the energy gain at least
matches the dissipative work which has to be done against the
pinning forces. The dissipative work per unit volume expended in
moving all GB by d$R$ is $\sigma_c b/(DR)$d$R$, and balancing
against the energy gain yields the limit grain size
\begin{equation}
R_l \approx \frac{\Gamma_0 D}{\sigma_c b} \approx \frac{\mu
b}{\sigma_c}\;.
\end{equation}
According to this relation, the grain size is inversely
proportional to the pinning stress. This gives a possibility to
experimentally assess the nature of the pinning by `tuning' the
pinning stress and measuring the grain size as a function of the
tuning parameters.

An obvious method to tune the pinning stress is to modify the
concentration of the pinning centers and to measure the impact
this has on grain size \cite{SMI-48}. However, this requires the comparison of
results from different samples and experimental results reported
in the literature are inconclusive \cite{MAN-98,OLG-86,MIO-00}. We
therefore refer to a quite different and unusual type of grain
growth experiment where the configuration of the pinning obstacles
is kept constant but the properties of the lattice are changed.

This type of grain growth experiment may be carried out on vortex
lattices of type-II superconductors in which quasi two-dimensional
grain structures are observed in (vortex polycrystals). In such a
system an external magnetic field penetrating the sample forms flux
lines that are disposed in a triangular lattice, whose elastic
properties (namely the lattice spacing and the elastic moduli) depend
on the magnetic field itself. A detailed theory of grain growth in
vortex polycrystals~\cite{MOR-03} can be developed along the same
lines followed here. In this case, it is worth pointing out that
the agreement of the theory with magnetic decoration data for the
average grain size is quite satisfactory, especially if compared to
the estimate based on local elasticity assumptions.

\section{Dynamics: Critical scaling near the depinning threshold}

Second-order phase transitions can be described by scaling laws and
critical exponents and the depinning transition is no exception. In
the system discussed here, the control parameter is the applied
stress, so that scaling laws depend on the distance from the critical
point $\sigma-\sigma_c$. In particular, as the system approaches the
transition the correlation length diverges as $\xi \sim
(\sigma-\sigma_c)^{-\nu}$. Similarly, one can define a characteristic
correlation time $t^*$, related to the correlation length as $t^* \sim
\xi^z$. The average dislocation velocity reaches a steady value,
scaling as $v \sim (\sigma-\sigma_c)^\beta$, above the transition, and
vanishes below. Before the steady-state the average velocity decays as
a power law $t^{-\theta}$, for times $t < t^*$. Furthermore, the
Orowan relation, a phenomenological law which relates the rate of
plastic deformation $\dot{\gamma}$ to average values of moving
dislocations density and velocity in a given crystal, implies that
similar scaling laws should hold for the strain rate
$\dot{\gamma}\equiv b\rho v$, where $\rho$ is the density of moving
dislocations, and $v$ their average velocity. In this respect, it is
thus tempting to establish a relationship between dynamical behavior
and the creep laws observed in plastically deforming crystals,
i.e. the crossover between primary (power law) to secondary (linear)
creep, due to their resemblance.

Scaling exponents also characterize the morphology of the dislocation
arrangement, which exhibits roughening close to the depinning
transition. The roughness can be quantified measuring the average
displacement correlations $C(x-x')=\langle (u(x)-u(x'))^2\rangle$. At
the transition in the steady-state, we expect a self-affine scaling
$C(x) \sim x^{2\zeta}$, where $\zeta$ is the roughness exponent while
the transient behavior is described by a scaling form of the type
$C(x,t)=t^{\beta_t} f(x/t^{1/z})$. As in ordinary critical phenomena,
only a fraction of the scaling exponents are independent. For
instance, one can easily derive the relations $\beta_t=z/\zeta$ and
$\theta=\beta /(\nu z)$.

In general, it has been shown that in the depinning problem there are
only two independent exponents that have been computed using the
renormalization group.  To connect our problem to previously obtained
results, we notice that the effective elastic energy of the pileup and
LAGB scales as $|q|$ in Fourier space, as in the problems of contact
line \cite{JOA-84} and planar crack depinning \cite{GAO-89}. We
can thus directly apply to our case the results obtained for a
manifold with long-range elastic energy \cite{KAR-98,ERT-94}. The
renormalization group analysis predicts that $d_c=3$ is the upper
critical dimension, above which fluctuations are suppressed. Thus for
$d>d_c$ there is no roughening (i.e.  $\zeta=0$) and the other
exponents can be computed in the mean-field approximation, yielding
$\beta=z=\nu=1$. These results are valid in the physically interesting
dimension $d=3$ apart from additional logarithmic corrections. For
$d<3$, a renormalization group expansion in $\epsilon = 3-d$ has been
performed to compute the exponents which at first order in $\epsilon$
are given by $\beta=7/9$, $\nu=3/2$, $\zeta=1/3$ and $z=7/9$ \cite{ERT-94}. Using
the scaling relation $\theta=\beta /(\nu z)$ one obtains $\theta=2/3$,
which coincides with the exponent of the so-called Andrade creep law,
observed in the creep deformation of several materials
\cite{Friedel,MIG-02}.

\subsection{Two-dimensional pileup: Numerical simulations}

In order to test the validity of the theoretical results, we have
performed a series of numerical simulations for the dynamics of a
two-dimensional pileup. This corresponds to an effective one
dimensional model in which $N$ interacting point dislocations move
along a line in presence of quenched disorder. For simplicity we
consider periodic boundary conditions, so that in absence of disorder
the equilibrium configuration is an equally spaced pileup. To test the
dependence on the system size, we change the dislocation number $N$
and the system size $L$ keeping the dislocation spacing $D=L/N$
constant.

\begin{figure}[tbh]
\centerline{\epsfig{file=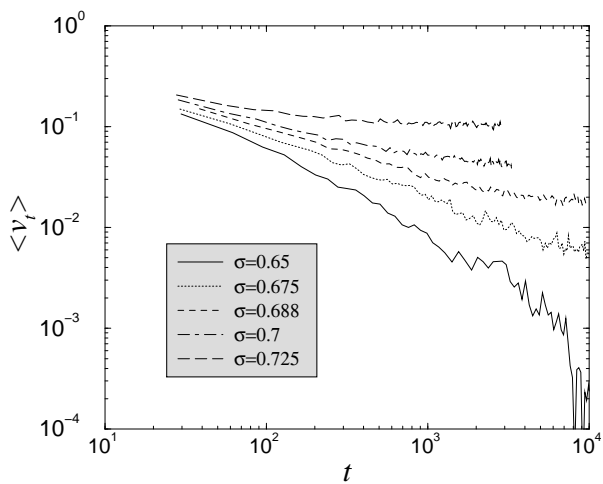,width=8cm,clip=!}}
\caption{The decay of the average pileup velocities as a function
of the applied stress $\sigma$. For $\sigma > \sigma_c \simeq
0.675$ the velocity reaches a steady value and decays to zero
otherwise.} \label{fig:vt}
\end{figure}

The equation of motion for the dislocation $i$ in the pileup is
given by
\begin{equation}
\chi \frac{d x_i}{dt} = \mu b^2 \sum_{j\neq i}
\frac{1}{|x_i-x_j|}+b\sigma + \sum_P f_p(x_i-X_P),
\label{eq:1dpileup}
\end{equation}
where $\chi$ is an effective viscosity and $\sigma$ is the applied
stress. The pinning centers are placed at randomly chosen positions
$X_P$ (with $P=1, ... N_P$) and exert an attractive force on the
dislocations
\begin{equation}
f_p(x)= -f_0 \frac{x}{\xi_p} e^{ -(x/\xi_p)^2}.
\end{equation}
In order to correctly take into account the effect of periodic
boundary conditions the interactions between dislocations are summed
over the images. In one dimension the sum can be performed exactly and
$1/|x|$ in Eq.~\ref{eq:1dpileup} is replaced by
\begin{equation}
\sum_{k=-\infty}^\infty \frac{1}{x+kL} =  \frac{\pi}{L \tan(\pi
x/L)}.
\end{equation}

The equation of motion (Eq.~\ref{eq:1dpileup}) is integrated
numerically using a Runge-Kutta algorithm for different values of the
applied stress. We take as initial condition a perfectly ordered
pileup, with equally spaced dislocations. For the simulations reported
here, we first considered $N=64,128,256,512$ dislocations with a
spacing $D=16$ and average pinning center spacing $d_p\equiv
L/N_p=2$. The units of time, space, and forces are chosen so that $\mu
b^2=1$, $\chi=1$, and $b=1$, and we set $f_0=1$ and $\xi_p=1$.

\begin{figure}[tbh]
\centerline{\epsfig{file=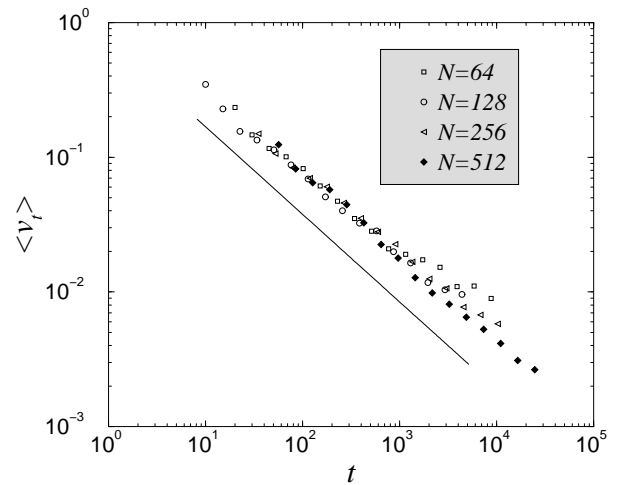,width=8cm,clip=!}}
\caption{The decay of the velocity at $\sigma > \sigma_c \simeq
0.675$ for different values of $N$. As $N$ increases the power law
scaling region extends. The line has a slope of $\theta = 0.65$.}
\label{fig:vt2}
\end{figure}

In Fig.~\ref{fig:vt} we report the time decay of the average pileup
velocity for different values of the applied stress. For large stress
values, $\sigma > \sigma_c \simeq 0.675$, the initial power law decay
is followed by a plateau, while the velocity decays to zero
otherwise. This allows to identify the depinning point as $\sigma_c
\simeq 0.675$. This is confirmed by the finite size analysis shown in
Fig.~\ref{fig:vt2}, indicating that for $\sigma_c = 0.675$ the power
law extends further as the system size is increased. The exponent of
the power law scaling $\theta \simeq 0.65$ is in good agreement with
the theoretical expectations.

\begin{figure}[tbh]
\centerline{\epsfig{file=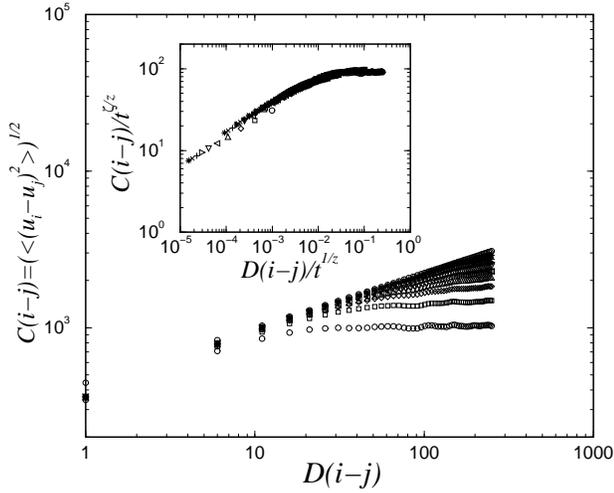,width=8cm,clip=!}}
\caption{The growth of the correlation function at the depinning
transition at different times. The data collapse in the inset
allows to estimate the roughness exponent $\zeta=0.35$ and the
dynamic exponent $z=0.9$.} \label{fig:corr}
\end{figure}

Moreover, in order to characterize the growth of correlations at the
critical point, we compute the displacement correlation function
$C(i-j,t)=(\langle (u_i(t)-u_j(t))^2\rangle)^{1/2}$ at different times
$t$ for $\sigma=\sigma_c$ (see Fig.~\ref{fig:corr}). The curves can be
collapsed using the scaling form $C(x,t)=t^{\zeta/z}f(x/t^{1/z})$ with
$\zeta=0.35$ and $z=0.9$ (see the inset of Fig.~\ref{fig:corr}). To
confirm this result we have also computed the evolution of the power
spectrum $P(k,t)=\int dx C(x) \exp (ikx)$. These curves can also be
collapsed as $P(k,t)=t^{(2\zeta+1)/z}g(kt^{1/z})$ with the same
exponent values as the correlation function.

\begin{figure}[tbh]
\centerline{\epsfig{file=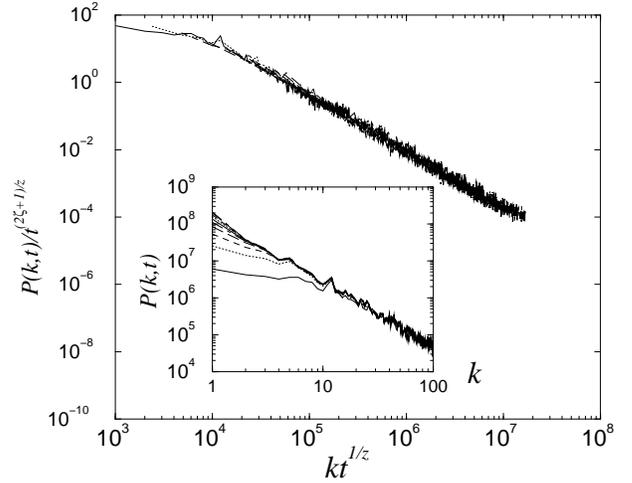,width=8cm,clip=!}}
\caption{The power spectrum of the pileup at the depinning
transition. The data collapse is consistent with the scaling of
the correlation function.} \label{fig:ps}
\end{figure}

In summary, all the exponents determined from the simulations are in
good agreement with the renormalization group predictions and with
previous simulations based directly on the elastic approximation,
confirming the validity of the elastic theory for the pileup.

\subsection{Three-dimensional pileup: relaxation of slip-band
growth rates}

Direct experimental observation of the velocity relaxation behavior of
planar dislocation arrangements is possible in certain alloys
exhibiting so-called planar slip where dislocations may form huge
pile-ups (see Figure~\ref{fig:pileup}). The motion of these planar
dislocation groups goes along with the formation of large slip steps
along the traces where the slip plane of the pileup intersects the
surface of the metal specimen. For a moving pile-up consisting of
roughly equally spaced dislocations, the slip step growth rate is
proportional to the dislocation velocity. Since the slip step growth
rate can be directly observed, this gives a possibility for a direct
experimental check of theoretical predictions.

\begin{figure}[tbh]
\centerline{\epsfig{file=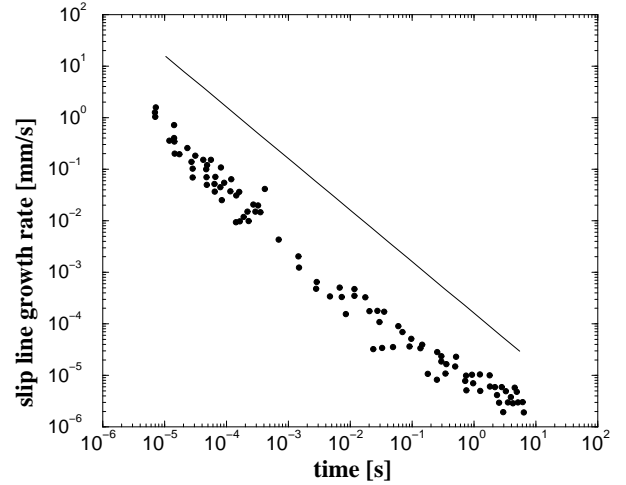,width=8cm,clip=!}}
\caption{Growth rate of slip steps on the surface of Cu-30at\% Zn
deformed at room temperature as a function of the time passed
after growth has started; after Ref.~\protect\cite{NEU-93}.
The line is a power law with exponent $\theta=1$.}
\label{fig:slipstep}
\end{figure}

Figure \ref{fig:slipstep} shows observed rates of slip step growth as
a function of the time after growth has started. The
double-logarithmic plot indicates relaxation of the growth rate (the
dislocation velocity) according to $v \propto t^{-\theta}$ with a
characteristic exponent $\theta = 1 \pm 0.1$ over six decades. This
indicates that the power-law is related to a depinning transition of
planar dislocation arrays in 3D for which we expect mean-field
exponents (see above) and according to the scaling relation
$\theta=\beta /(\nu z)$ a value $\theta = 1$. The apparent length of
the scaling regime indicates that driving of the dislocation arrays
occurs at stresses very close to the critical one. This is in line
with the general observation that dislocation arrangements in slowly
deforming crystals (where 'slow' covers the entire range of strain
rates used in typical experiments, \cite{ZAI-01}) are in a
close-to-critical state \cite{ZAI-01,MIG-01}.

\section{Conclusions}

We have investigated the depinning transition of planar dislocation
arrays such as small-angle grain boundaries or dislocation
pile-ups. Contrary to the case of isolated dislocations, the elastic
interactions between dislocation line segments in such arrays are of
long-range nature and, hence, cannot be described within a line- or
surface-tension approximation. The pinning of planar dislocation
arrays has been investigated both in the weak and strong pinning
limits using collective pinning theory and Friedel statistics,
respectively.  The results have been applied to the problem of grain
growth limited by grain boundary pinning referred to as Zener pinning.

Long-range elastic interactions also govern the dynamics of planar
dislocation arrays at the depinning threshold. In two dimensions,
computer simulations and theoretical arguments suggest that the
dynamics falls into the same class as contact-line depinning,
while in three dimensions the dynamical behavior can be described
by mean-field exponents. In particular, we have demonstrated that
the mean-field prediction for the velocity relaxation of a planar
dislocation array (pile-up) is consistent with experimental
observations of the time-dependent growth of slip bands in alloys
exhibiting planar slip.

The dislocation arrangements discussed in the present study have a
simple, quasi-planar geometry in which only dislocations of one sign
are present and only small perturbations of the planar arrangement of
the dislocations are permitted. Because of this particular geometry,
the dislocation assemblies behave like two or three- dimensional
long-range elastic manifolds. The situation is much more complicated
when dislocations of different types and directions of motion have to
be considered. In such situations, there is still a transition between
a stationary and a moving state of the dislocation assembly ('yielding
transition').  However, in general dislocation assemblies the
existence of metastable stationary states does not depend on the
presence of quenched disorder as in the present study. Rather, the
interactions between dislocation lines of different type together with
the dynamics constraints which tie the motion of the dislocation lines
to their respective slip planes lead to the possibility of forming
metastable jammed configurations even in the absence of any
disorder. While the general scenario of dynamic non-equilibrium phase
transitions applies to such systems, no ready-made theoretical
framework is available and, hence, a theory of the yielding and
dynamic behavior of general dislocation systems remains a formidable
task for future investigations.


\begin{thebibliography}{99}
\bibitem{Footnote1}
Because of the emergence of `corners' in the dislocation lines,
this yields a rather poor description of the interaction between
adjacent line segments. However, in the following we are mainly
interested in the long-wavelength elasticity of the dislocation
lines, which in this staircase approximation is captured exactly
in the limit of infinitesimally small segment lengths.
\end{thebibliography}
\end{document}